# Bulk and interface spin-orbit torques in Pt/Co/MgO thin film structures

M.S. Gabor[1,*], M. Belmeguenai[2], I.M. Miron[3]

[1]*Center for Superconductivity, Spintronics and Surface Science, Physics and Chemistry Department, Technical University of Cluj-Napoca, Str. Memorandumului, 400114 Cluj-Napoca, Romania*
[2]*Université Sorbonne Paris Nord, LSPM, CNRS, UPR 3407, F-93430 Villetaneuse, France*
[3]*Université Grenoble Alpes, CNRS, CEA, Grenoble INP, SPINTEC, Grenoble, France*
[*]mihai.gabor@phys.utcluj.ro

*Abstract*

We investigate the origin of spin-orbit torques (SOTs) in archetypical Pt/Co/MgO thin films structures by performing harmonic Hall measurements. The behaviour of the damping like (DL) effective field ($h_{DL}$) with varying the Pt layer thickness and the Co layer thickness indicates that bulk spin-Hall effect (SHE) in Pt is mainly responsible for DL-SOT. The insertion of a Pd ultrathin layer at the Pt/Co interface leads to a step decrease in $h_{DL}$, attributed to the modification of interfacial spin transparency. Further increase in Pd thickness led to a reduction of the interfacial spin-orbit coupling (iSOC) quantified by the decrease in the surface magnetic anisotropy. The consistent insensitivity of $h_{DL}$ to variations in iSOC at the bottom Pt/Co interface and oxidation at the top Co/MgO interface provides additional evidence for the bulk SHE origin of DL-SOT. The strong reduction in the field-like (FL) torque effective field ($h_{FL}$) with decreasing iSOC at the Pt/Co interface points to the interfacial nature of FL-SOT, either due to iSOC induced interfacial spin-currents or to the Rashba-Edelstein effect at the Pt/Co interface. Furthermore, we demonstrate that a FL-SOT develops at the top Co/MgO interface opposing the one generated at the bottom Pt/Co interface, whose strength increases with Co/MgO interfacial oxidation, and attributed to the Rashba-Edelstein effect.

*Introduction*

The current induced spin–orbit torques (SOTs) in heavy-metal (HM)/ferromagnet (FM) heterostructures [1,2] have garnered remarkable research interest for the development of electrically controlled spintronic and spin-logic devices [3-9]. Two types of mechanisms are generally considered for the microscopic origin of SOTs: either bulk or interface related. In one case, the spin-Hall effect (SHE) [10,11] due to the spin-orbit coupling (SOC) in the bulk of the HM layer produces a spin current propagating towards the HM/FM interface. There, it is partially or totally absorbed by the FM layer as torques on the magnetization. The other mechanism, inverse spin galvanic effect (IGSE) or the Rashba-Edelstein effect (REE), occurring at interfaces with broken inversion symmetry, involves an in-plane charge current generating a spin accumulation via interface spin-orbit-coupling (iSOC), ultimately exerting torques on the magnetization of the FM layer through exchange coupling [12]. Recently, other interfacial mechanism based on iSOC were proposed to generate SOTs at the HM/FM interfaces [13-15]. Both the bulk and interfacial mechanisms are expected to produce two types of torques on the magnetization with different symmetries: damping-like (DL) and field-like (FL) [11,16-19].

From an applications perspective, it is of major importance to disentangle the origin of SOTs for a particular HM/FM structure to facilitate their optimization. Given that both bulk and interface mechanisms can coexist within the same samples, unravelling the exact nature of the SOTs is not a straightforward experimental task. Varying the thickness of the HM layer to test the SHE as a possible origin for SOTs influences the electrical resistivity of the HM, which, in turn, affects the generation of spin current via the SHE [20]. It could also influence the strains in the HM/FM or even the interfacial morphology, which would impact the SOTs [21]. On the other hand, engineering the interfaces to modify the iSOC might also modify the SOTs beyond the interface-related mechanism. It could strongly affect the current distribution within the stack or it could impact the spin memory loss (SML) at the HM/FM interface and influence the SOTs generated by the bulk SHE [22,23].

In this paper, we investigate the nature of the SOTs in the archetypical Pt/Co/MgO thin films structure. In this type of structure three possible mechanisms could produce both DL and FL SOTs on the FM layer magnetization: (i) SHE in Pt, (ii) REE and/or interfacial spin-currents induced by iSOC at the Pt/Co interface, and (iii) REE at the top Co/MgO interface. Initially, we explore the dependence of the DL-SOT and FL-SOT on the Pt layer thickness, revealing that SHE is the main source of DL-SOT [mechanism (i)]. Upon varying the Co layer thickness, DL-SOT behaves as expected for SHE, while FL-SOT deviates from the expected behaviour, suggesting the influence of the other interfacial mechanisms beyond the SHE.

Furthermore, we engineer the Pt/Co interface by the insertion of the ultrathin Pd layer to tune the iSOC. Interestingly, DL-SOT does not scale with iSOC, while the FL-SOT strongly correlates with it, indicating that the REE-like interfacial mechanism (ii) is dominant for FL-SOT. Lastly, by adjusting the oxidation level at the Pt/Co interface through changes in the MgO layer thickness we find that DL-SOT remains unaffected, whereas FL-SOT scales with the Co/MgO surface magnetic anisotropy, which is a measure of the interfacial charge transfer affecting the Rashba field at this interface. These findings suggest that REE mechanism (iii) does not impact the DL-SOT, even though it is instrumental in generating the FL-SOT.

*Experimental*

All the samples studied here were grown at room temperature on thermally oxidized silicon substrates using an ultrahigh vacuum system that integrates electron beam evaporation and magnetron sputtering. The typical sample configuration is as follows: Si/SiO$_2$//Ta (2)/Pt ($t_{Pt}$)/Co ($t_{Co}$)/MgO (2)/Ta (1.5), with the values in parentheses indicating the thicknesses in nanometers. Additional samples were also grown, and their structural details will be addressed later in the text. The 2-nanometer-thick tantalum (Ta) seed layer was deposited using direct current (dc) sputtering onto the substrate under an argon pressure of 1 mTorr. Subsequently, the argon gas was purged from the system, and the remaining structure was deposited through electron beam evaporation. Throughout the deposition of metallic layers, the chamber pressure remained within the $10^{-10}$ Torr range, whereas during the evaporation of MgO, the pressure increased to around $10^{-8}$ Torr. To protect the structure from contamination due to exposure to the atmosphere, a 1.5-nanometer-thick Ta capping layer was dc sputtered on the substrate under an argon pressure of 1 mTorr. For the fabrication of the active part of the samples we used the electron beam evaporation technique due to its ability to generate well-defined interfaces compared to sputtering. Moreover, the directional nature of the incoming atomic flux allowed for the deposition of wedge-shaped layers using a movable shutter placed in front of the substrate. The specificity of the wedge ensured the simultaneous deposition of each series of samples, thereby eliminating material variations that might arise in sequential deposition runs.

The saturation magnetization of the samples was measured at room temperature using a vibrating sample magnetometer (VSM). For magneto-electric experiments, the samples were patterned through conventional UV photolithography and argon-ion milling techniques. A dual photoresist process was employed to create an undercut in the photoresist mask, reducing edge roughness and re-deposition during the milling process. Electrical resistance measurements were performed using the standard four-point technique, while the evaluation of Spin-Orbit Torques (SOTs) was carried out using the harmonic Hall voltage technique [24-27].

*Results and discussions*

One of the experimental approaches used to investigate the bulk and interfacial characteristics of SOTs involves studying their dependence on the thickness of the heavy metal Pt layer ($t_{Pt}$). In principle, effects arising from the REE or other interfacial interactions should remain independent of $t_{Pt}$, whereas effects due to the bulk SHE would depend on $t_{Pt}$, for thicknesses of the order of the spin diffusion length. Thus, the thickness dependence of the torques should provide information about their physical origin. However, it is important that the overall structure and morphology (crystallinity of the layers, interfacial roughness, interdiffusion processes) remain as much as possible unchanged upon modifying $t_{Pt}$. This helps to minimize other factors influencing the SOTs. To this end, we deposited a Si/SiO$_2$//Ta (2)/Pt (1.5-5)/Co (2)/MgO (2)/Ta (1.5) sample stack in which the Pt film was grown as a wedge layer over 20 mm. A series of $50 \times 10$ $\mu m$ Hall crosses with increasing $t_{Pt}$ were patterned on the substrate along the wedge direction and diced for magnetoelectric measurements. Figure 1(a) shows a schematic representation of the SOTs harmonic Hall voltage measurement geometry. Since all the samples are in-pane magnetized, we used a variant of the harmonic Hall method adapted for such samples by Avci *et al*. [28], which provides a straightforward method of excluding the thermo-electric effects. The technique involves injecting an AC current ($I_\omega = I \sin \omega t$) into the patterned stripe (along $\hat{x}$) and measuring the first ($R_\omega = V_\omega/I$) and the second ($R_{2\omega} = V_{2\omega}/I$) harmonic Hall resistances (along $\hat{y}$), while rotating the magnetization in-plane by applying an external in-plane rotating magnetic field (*H*). The *R$_\omega$* provides information about the planar Hall effect as [28] $R_\omega = R_{PHE} \sin 2\varphi_H$, where $R_{PHE}$ is the planar Hall resistance and $\varphi$ is the azimuthal angle of the magnetization from the current direction [Fig. 1(b)]. The samples have a relatively weak in-plane uniaxial anisotropy; thus, the magnetization follows the external applied magnetic field and the $\varphi$ azimuthal angle of the magnetization practically duplicates the $\varphi_H$ azimuthal angle of the field. The second harmonic Hall resistance *R$_{2\omega}$* contains information about the SOT effective fields and it is given by [28]

$$R_{2\omega} = \frac{1}{2}\left(R_{AHE}\frac{h_{DL}}{H+H_k} + R_{\nabla T}\right)\cos\varphi + R_{PHE}(2\cos^3\varphi_H - \cos\varphi_H)\frac{h_{FL}+h_{Oe}}{H},$$

where $R_{PHE}$ and $R_{AHE}$ are the planar and anomalous Hall resistances, $h_{DL}$ and $h_{FL}$ are the damping-like and field-like effective fields, $H_k$ is the perpendicular anisotropy field, $R_{\nabla T}$ is the second harmonic Hall resistance due to thermo-electric effects and $h_{Oe}$ is the Oersted field produced by the charge current passing through the HM layer. $R_{AHE}$ and $H_k$ are determined by applying an out-of-plane field and measuring the transverse voltage ($V_{xy}$) which is then divided by the electrical current (*I*) passing through the device to give the transverse resistance $R_{xy} = V_{xy}/I$. The $R_{AHE}$ is calculated as $[R_{xy}(+M_z) - R_{xy}(-M_z)]/2$, where $R_{xy}(+/-M_z)$ is the transverse resistance for positive/negative saturation. The inset of Fig. 1(c) shows a representative AHE resistance measurement where the $R_{AHE}$ and the $\mu_0 H_k$ are indicated. By fitting the $R_{2\omega}$ experimental data to the above equation, two contributions can be extracted: one which shows a $\cos\varphi_H$ dependence and gives information about $h_{DL}$, and another which shows a $2cos^3\varphi_H - \cos\varphi_H$ dependence and gives information about $h_{FL}$ [Fig. 1(b)]. The $h_{DL}$ and the $R_{\nabla T}$ are obtained from the slope and the intercept of the linear fit of the $\cos\varphi_H$ contribution dependence on the inverse of the sum of the external and anisotropy fields [Fig. 1(c)]. The sum $h_{FL} + h_{Oe}$ is determined from the slope of the linear fit of the $2cos^3\varphi_H - \cos\varphi_H$ contribution dependence on the inverse external field [Fig. 1(d)]. The Oersted field is calculated as $h_{Oe} = \mu_0 j_{Pt} t_{Pt}/2$, where $j_{Pt}$ is the charge current density through the Pt layer, and it is subtracted from $h_{FL} + h_{Oe}$ sum to obtain $h_{FL}$. The $j_{Pt}$ was calculated by assuming a parallel resistor model and by subtracting the contribution to the total resistance of the Ta (2)/Co (2)/MgO (2)/Ta (1.5) stack, measured on a sample deposited on the same run (see Supplemental Material [29]).

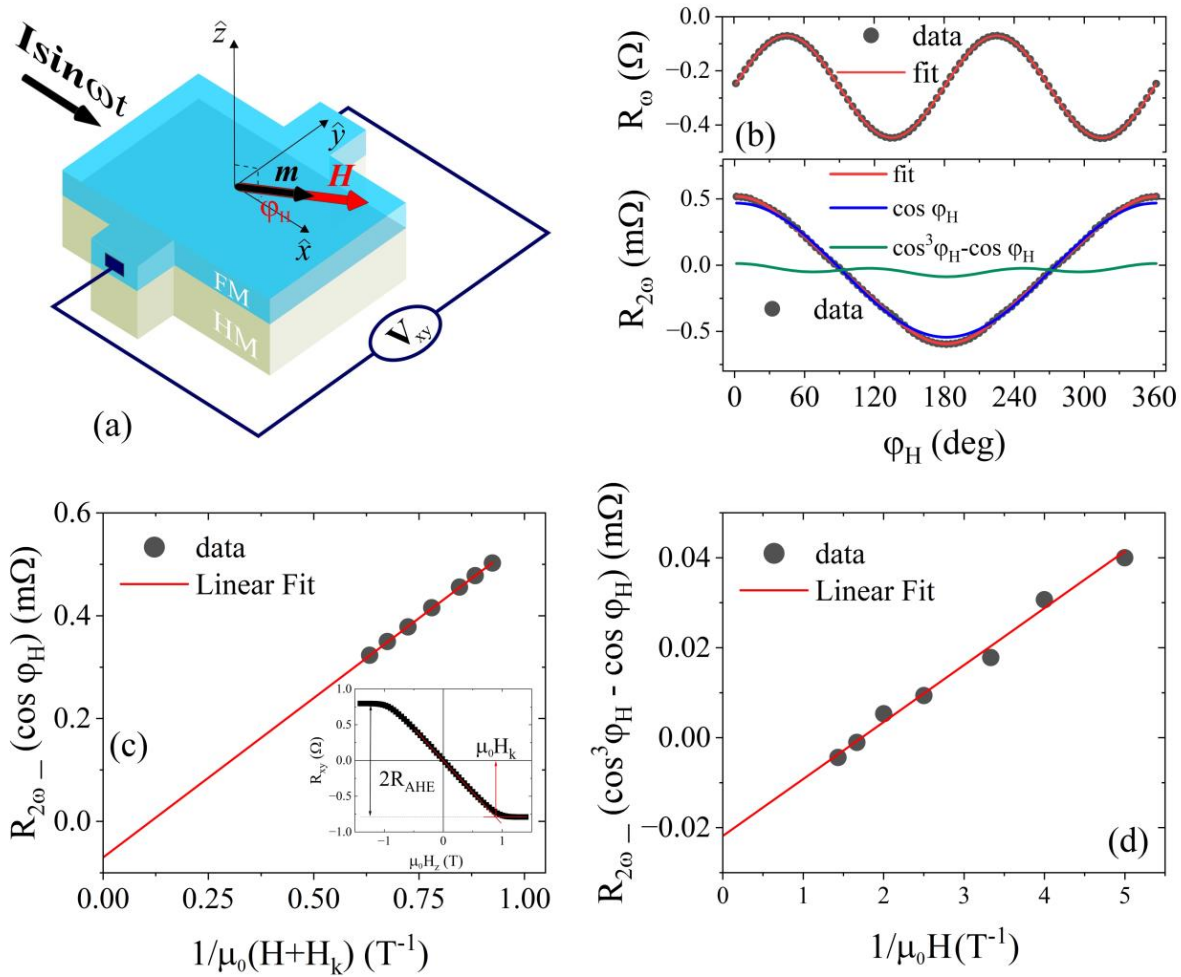

FIG. 1. (a) Schematic representation of the harmonic Hall magneto-transport measurement geometry. (b) First and second harmonic Hall resistances, as a function of the azimuthal angle of the external in-plane magnetic field with the current direction, for the Pt (4.68)/Co (2)/MgO (2) representative sample. The points are experimental data while the continuous lines are fits using the equations from the main text. Dependence of the second harmonic Hall resistance (c) $\cos\varphi_H$ – contribution on $1/\mu_0(H + H_k)$ and (d) $2cos^3\varphi_H - \cos\varphi_H$ – contribution on $1/\mu_0 H$, used to extract the $h_{DL}$ and the $h_{FL}$. The straight lines are linear fits to the data. The inset in (c) shows the transverse resistance as a function of the perpendicular applied field, used to extract $R_{AHE}$ and $\mu_0 H_k$.

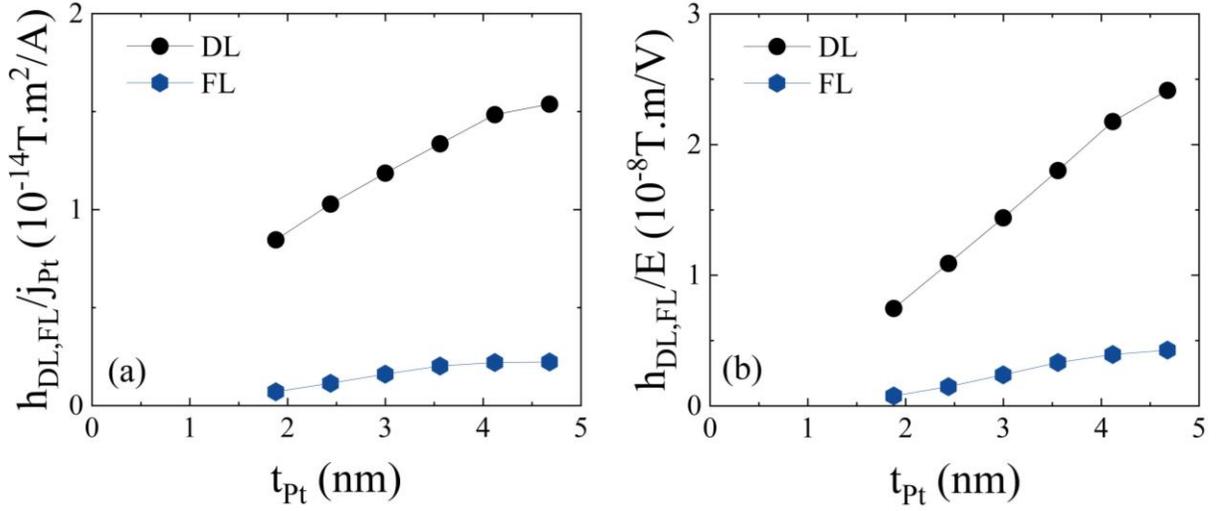

FIG. 2. Damping-like ($h_{DL}$) and field-like ($h_{FL}$) effective fields normalized by the (a) charge current density through the Pt layer ($j_{Pt}$) and by (b) the applied electric field ($E = \rho_{Pt} j_{Pt}$), as a function of the Pt layer thickness ($t_{Pt}$).

Figure 2 shows the damping-like ($h_{DL}/j_{Pt}$, $h_{DL}/E$) and field-like ($h_{FL}/j_{Pt}$, $h_{FL}/E$) effective fields normalized by either the charge current density through the Pt layer or by the applied electric field ($E = \rho_{Pt} j_{Pt}$). Irrespective of the normalization procedure, the effective fields show an increase with increasing $t_{Pt}$, with a tendency for saturation at larger $t_{Pt}$. Both the values and the behaviour are in agreement with literature [20]. Moreover, this type of effective fields HM thickness dependence is a recurring feature in various systems that rely on different HM layers [30-34]. Our findings are consistent with first-principles calculations based on the drift-diffusion formalism of the SHE. These calculations suggest that both damping-like and field-like torques should exhibit a similar dependence on the thickness of the heavy metal (HM) layer and that the damping-like torque should be significantly larger than the field-like torque given their direct proportionality to the real and imaginary parts of the spin mixing conductance[16]. Thus, our results seem to indicate that the SHE is the main source of the spin current responsible for the two types of torques. Nonetheless, caution should be exercised in drawing a conclusive statement. REE-like interfacial mechanisms generating the torques might be dominated by the SHE for relatively large HM thicknesses. Therefore, even smaller thicknesses of the HM layer should be probed to evidence interfacial effects [16,17]. However, further reducing the thickness of the HM layer may bring additional complications. The resistivity of the Pt layer increases strongly with degreasing thickness [29] which affects the spin current generation via SHE [20] and also the charge current distribution within the stack. Also, first-principles calculations showed that the strains, which are expected to increase with decreasing the HM thickness, have a strong impact on both damping-like and field-like torques [35]. Moreover, reducing the thickness could influence the HM/FM interfacial morphology affecting the spin-current transmission across the interface or interfacial spin-current generation.

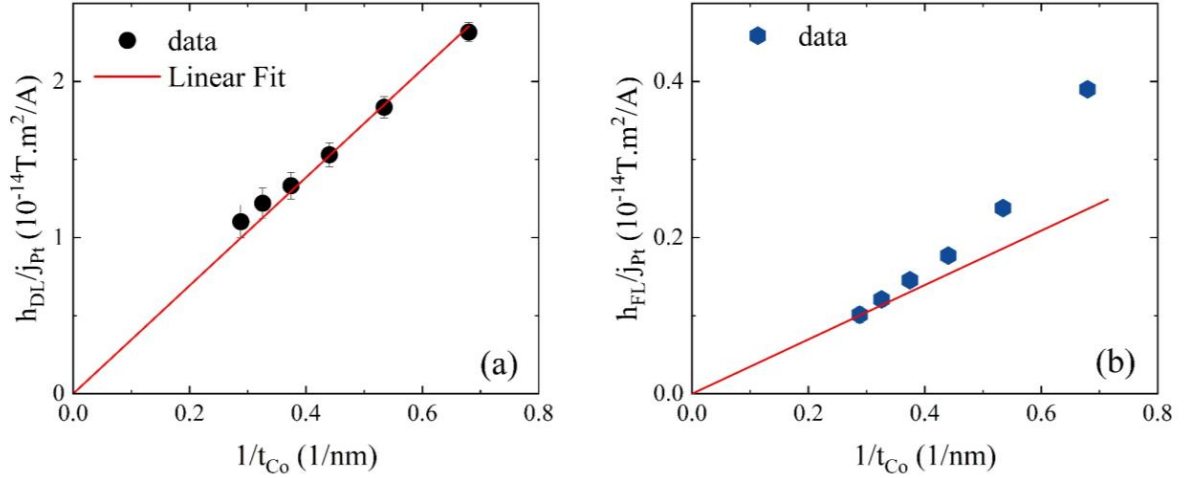

FIG. 3. (a) Damping-like ($h_{DL}/j_{Pt}$) and (b) field-like ($h_{FL}/j_{Pt}$) effective fields normalized by the charge current density through the Pt layer ($j_{Pt}$) as a function of the inverse Co layer thickness ($1/t_{Co}$). The line in (a) is a linear fit, while in (b) the line is drawn from the origin through the first data point.

Supplementary, one can examine the torques dependence on the thickness of the ferromagnetic layer ($t_{Co}$). If the bulk SHE within the Pt layer is the source of the torques, and as a result, the source of the torques lies outside the Co layer, one would expect the effective fields to be inversely proportional to the thickness of the Co layer ($\propto 1/t_{Co}$) [16]. To study this, we fabricated a series of samples with the structure Si/SiO$_2$//Ta (2)/Pt (5)/Co (1.4-4)/MgO (2)/Ta (1.5), maintaining a constant Pt layer thickness while varying the Co layer thickness in a wedge-like manner. Figure 3 shows the $h_{DL}/j_{Pt}$ and $h_{FL}/j_{Pt}$ as a function of the inverse Co layer thickness ($1/t_{Co}$). From Fig. 3(a) one can observe that $h_{DL}/j_{Pt}$ decreases linearly with $1/t_{Co}$, indicating that the bulk SHE is the main source of $h_{DL}$. As seen in Fig. 3(b), $h_{FL}/j_{Pt}$ deviates from the linearity and does not corelate with $h_{DL}/j_{Pt}$. This suggests the presence of an additional interfacial mechanisms contributing to the $h_{FL}$. It is to be mentioned that a deviation from the perfect $1/t_{Co}$ dependence is expected as long as the *interfacial layers* where spin-current generation or spin accumulation takes places have a finite thickness [12,36-38]. A similar behaviour is expected for the SHE induced $h_{FL}$ when the thickness of the ferromagnetic layer is below the spin decoherence length [39]. However, this does not apply to our samples because the thickness of the Co layer exceeds the spin decoherence length for Co, which is approximately 1.2 nm [40]. We can also rule out the effect of strains in the Co layer, at least for thicknesses larger than 1.5 nm. Magnetic anisotropy measurements [29], which are highly sensitive to strains in Co [41], did not indicate the presence of significant strains.

Interface effects influencing the SOTs could be associated with either the Pt/Co or Co/MgO interfaces, or possibly both. First, we will consider the bottom Pt/Co interface. REE or other interfacial related mechanisms generating torques rely on the interfacial spin orbit coupling (iSOC) [13,16], thus tuning the iSOC could provide insight into the interfacial generation of the SOTs. Our strategy to tune the iSOC is to insert an ultrathin Pd layer at the interface between Pt and Co. Hence, we fabricated a series of samples with the structure Si/SiO$_2$//Ta (2)/Pt (5)/Pd (0-1.8)/Co (2)/MgO (2)/Ta (1.5), maintaining Pt and Co layers thicknesses constant, while varying the Pd layer thickness in a wedge-like manner. We selected Pd as an interlayer for several reasons. Pt and Pd share similar crystal structures, both belonging to the Fm-3m space group, with closely matched lattice parameters ($a_{Pt} = 0.392$ nm and $a_{Pd} = 0.389$ nm), which facilitates the high-quality layer-by-layer growth of Pd on Pt. They also exhibit comparable bulk electrical resistivities ($\rho_{Pt} = 106$ nΩm and $\rho_{Pd} = 105$ nΩm) ensuring a uniform current flow in the Pt/Pd bilayer [29]. Moreover, Pd has a lower SOC than Pt and a relatively large spin-diffusion length ($\lambda_{sd} \sim 8$ nm) [21].

Figure 4(a) shows the damping-like ($h_{DL}/j_{Pt,Pd}$) effective field normalized by the charge current density through the Pt/Pd bilayer ($j_{Pt,Pd}$) as a function of the thickness of the Pd layer ($t_{Pd}$). The $h_{DL}/j_{Pt,Pd}$ does not show a clear dependence on $t_{Pd}$, it remains constant within the error bars for the entire $t_{Pd}$ range. It is noteworthy to observe that that the insertion of the Pd layer leads to an around 5% decrease in $h_{DL}/j_{Pt,Pd}$ compared to the Pt/Co/MgO sample without the Pd insertion layer. The field-like effective field ($h_{FL}/j_{Pt,Pd}$) normalized by $j_{Pt,Pd}$ is shown in Fig. 4(b) and displays a different behaviour. It exhibits a continuous decrease with $t_{Pd}$ up to approximately a 1 nm thickness, after which it remains relatively constant. It is also interesting to observe that $h_{FL}/j_{Pt,Pd}$ changes sign for $t_{Pd}$ larger than approximately 0.8 nm. For clarity, in Fig. 4(b) the measured transverse effective field and the Oe field, extracted from the measured transverse effective field to determine the $h_{FL}$, are also displayed.

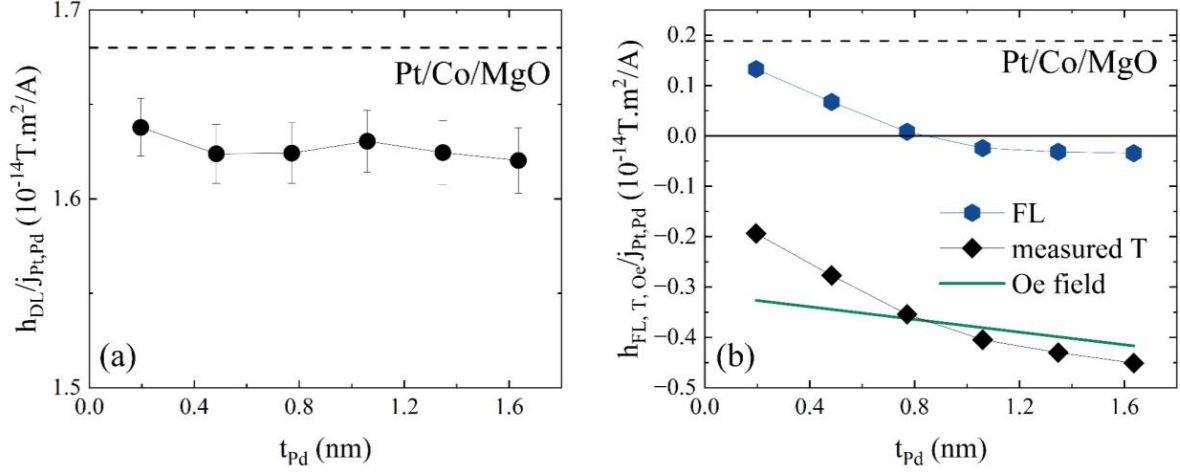

FIG. 4. (a) Damping-like ($h_{DL}/j_{Pt,Pd}$) and (b) field-like ($h_{FL}/j_{Pt,Pd}$) effective fields normalized by the charge current density through the Pt/Pd bilayer ($j_{Pt,Pd}$) as a function of the thickness of the Pd layer. The measured transverse effective field (measured T) and the Oersted field, extracted from the measured transverse effective field to determine the $h_{FL}$, are also displayed. The dotted lines indicate the corresponding effective fields values for the samples without the interfacial Pd layer.

We recently showed that the insertion of an ultrathin Pd layer at the Pt/Co interface can effectively screen the SOC of Pt [42]. Moreover, it is well known that the surface magnetic anisotropy ($K_s$) at the HM/Co interface is related to the iSOC enhanced interface orbital magnetic moments [43]. Hence, the variation of iSOC with respect to $t_{Pd}$ could be assessed by measuring the dependence of $K_s$ on $t_{Pd}$ [29]. Figure 5 shows the $h_{DL}/j_{Pt,Pd}$ and $h_{FL}/j_{Pt,Pd}$ as a function of $K_s$. Interestingly, $h_{DL}/j_{Pt,Pd}$ does not scale with $K_s$, while $h_{FL}/j_{Pt,Pd}$ shows a linear correlation with $K_s$.

We will start by examining the behaviour of $h_{DL}/j_{Pt,Pd}$. The reduction of $h_{DL}/j_{Pt,Pd}$ with the insertion of the Pd layer relative to the Pt/Co/MgO sample can be understood by considering the various mechanisms involving SHE and interface-generated spin currents, or a combination of both. This decrease could be attributed to the spin memory loss (SML) at the interface [22,44], which reduces the SHE generated spin-current transmission through the interface [45]. However, this effect is unlikely as it involves the loss of spin information due to spin-flip scattering, which should increase with iSOC [23] and, therefore, with $K_s$. This is contrary to our observation that the decrease is independent of $K_s$ [Fig. 5(a)]. Another possibility is that the Pd layer produces a spin-current via SHE opposing the one produced by the Pt layer. This is also not likely having in view the large spin diffusion length of Pd relative to $t_{Pd}$ and the relatively low spin-Hall angle of Pd [21,34]. Another possibility is the generation of spin currents at the Pd/Co interface producing a $h_{DL}$ opposing the one produced via the bulk SHE in Pt. Theoretical calculations indicated that spin-currents could be generated by interfaces when iSOC is present [17,46], or even at the interface between a low SOC normal metal and ferromagnetic material [13]. Although we cannot exclude this possibility, it is not likely to be the case since one would expect a variation of $h_{DL}$ with iSOC, and thus $K_s$. Another plausible mechanism, consistent with our observations and independent of iSOC, is the modification of the spin transparency at the Pt/Co interface by introducing the Pd layer. This is an electronic effect concerning the transmission and reflection of electrons carrying angular momentum. It is linked to the electronic band matching of the two metals across the interface and is not associated with the loss of spin polarization [47].

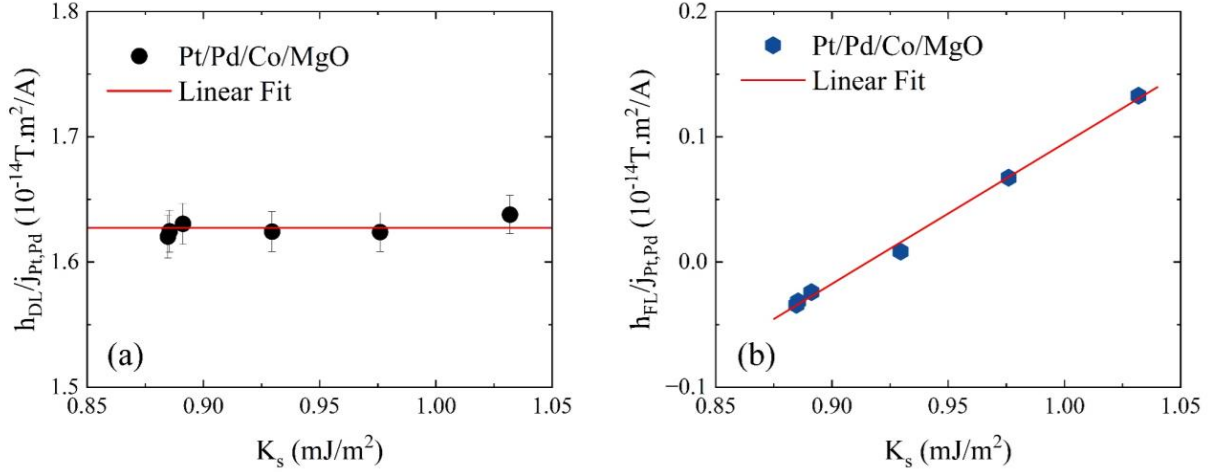

FIG. 5. (a) Damping-like ($h_{DL}/j_{Pt,Pd}$) and (b) field-like ($h_{FL}/j_{Pt,Pd}$) effective fields normalized by the charge current density through the Pt-Pd layer ($j_{Pt,Pd}$) as a function of the surface magnetic anisotropy ($K_s$). The lines are linear fits to the data.

The behaviour of $h_{FL}/j_{Pt,Pd}$ is quite different. Besides the similar initial reduction with the insertion of the thinnest Pd layer, $h_{FL}/j_{Pt,Pd}$ scales linearly with $K_s$ and, thus, with iSOC. This indicate that besides SHE there is another interfacial mechanism responsible for $h_{FL}$. The reduction of $h_{FL}$ by the insertion of the Pd layer [Fig.4(b)] and the linear correlation $K_s$ [Fig.5(b)] indicates that the primary mechanism responsible for generating $h_{FL}$ is of an interfacial nature. This could be attributed to either REE-like mechanism [16,37,48] or interfacial spin-currents induced by iSOC [13,17], which result in a net interfacial spin accumulation responsible for $h_{FL}$. One might argue that the insertion of the Pd layer diminishes the proximity-induced magnetization (PIM) in Pt, providing a potential mechanism for altering $h_{FL}$ through the dephasing of the spin accumulation by the PIM exchange field, as an alternative to the variation of the iSOC. However, it was shown that PIM in Pd/Co is only marginally lower than in Pt/Co[49]. Furthermore, the decrease of the PIM is expected to increase $h_{FL}$ and not diminish it [50].

As we will demonstrate in the following sections, the observed sign change of $h_{FL}/j_{Pt,Pd}$ for $t_{Pd}$ larger than about 0.8 nm can be attributed to the emergence of a FL-SOT at the top Co/MgO interface, that possesses an opposite sign compared to the one at the bottom Pt/Co interface. The reason $h_{FL}/j_{Pt,Pd}$ remains constant with further increase of $t_{Pd}$ is because roughly 1 nm of Pd is sufficient to screen the SOC of Pt [42]. Further increasing the Pd layer thickness will not further reduce the iSOC. It will remain constant and associated with the iSOC of the Pd/Co interface, which although smaller than the one related to the Pt/Co interface, it is not negligible.

To summarize, our data indicates that $h_{DL}$ is primarily generated via the bulk SHE from Pt, and the decrease in $h_{DL}$ following the insertion of the Pd layer can be attributed to the variation of the interfacial spin transparency. At the same time, besides SHE, the results suggests that the primary mechanism responsible for $h_{FL}$ is of interfacial nature related either to iSOC induced interfacial spin-currents or to REE at the Pt/Co interface. Since the current flow through the Pt/Pd bilayer is uniform [29], the linear relationship between $h_{FL}$ and $K_s$ will hold, even when $h_{FL}$ is normalized by the charge current density through the Pd interlayer ($h_{FL}/j_{Pd}$). This supports our observation that the primary mechanism responsible for $h_{FL}$ is of interfacial nature.

Up to this point, we focused on examining the impact of the bottom Pt/Co interface on the SOTs. However, it is also reasonable to consider that the top Co/MgO interface may play a significant role in SOTs generation. For this purpose, we deposited a series of Si/SiO$_2$//Ta (2)/Pt (5)/Co (2)/MgO (0-3.6)/Pt (5) samples, where the MgO was grown as a wedge layer with thicknesses ranging up to 3.6 nm. The Ta (1.5) capping layer was replaced with a Pt (5) film for two main reasons. It will allow us to study the impact on SOTs upon continuously separating the Co/Pt interface. Also, when the MgO layer is sufficiently thick, the top Pt layer is expected to generate only an Oersted field countering the one produced by the bottom Pt layer, thus making the determination of $h_{FL}$ more reliable [29].

Figure 6(a) shows the electrical resistance ($R_{xx}$) of the stacks as a function of the MgO layer thickness ($t_{MgO}$). Initially, there is a relatively strong increase in $R_{xx}$ with $t_{MgO}$. This increase is most likely attributed to the discontinuity of the MgO layer within this thickness range, which increases interface scatterings and subsequently leads to a higher $R_{xx}$. As $t_{MgO}$ increases, the layer becomes continuous and $R_{xx}$ starts to drop. Interestingly, above a MgO layer thickness corresponding to 3-4 atomic planes, the $R_{xx}$ falls below that of the sample with no MgO layer. In the case of the Pt/Co/Pt sample, it is probable that there is some degree of intermixing at the Co/Pt interface, leading to increased interface scatterings. The introduction of the MgO layer serves to prevent this intermixing, resulting in a decrease in $R_{xx}$. However, for larger MgO thicknesses, the $R_{xx}$ exhibits a slight increase of approximately 2%, which we attribute to a minor oxidation of the Co layer with increasing MgO layer thickness. This small variation of $R_{xx}$ for $t_{MgO}$ above 0.6-0.7 nm, ensures that the current distribution remains relatively unchanged upon increasing $t_{MgO}$ up to 3.6 nm. The behaviour of the $R_{AHE}$ as a function of $t_{MgO}$, shown in the inset of Fig. 6(a), is consistent with that of $R_{xx}$. It initially decreases, after which it remains relatively independent of $t_{MgO}$. With the insertion of the MgO layer, the Co/Pt interface contribution to AHE is eliminated, resulting in the initial decrease of $R_{AHE}$ with $t_{MgO}$.

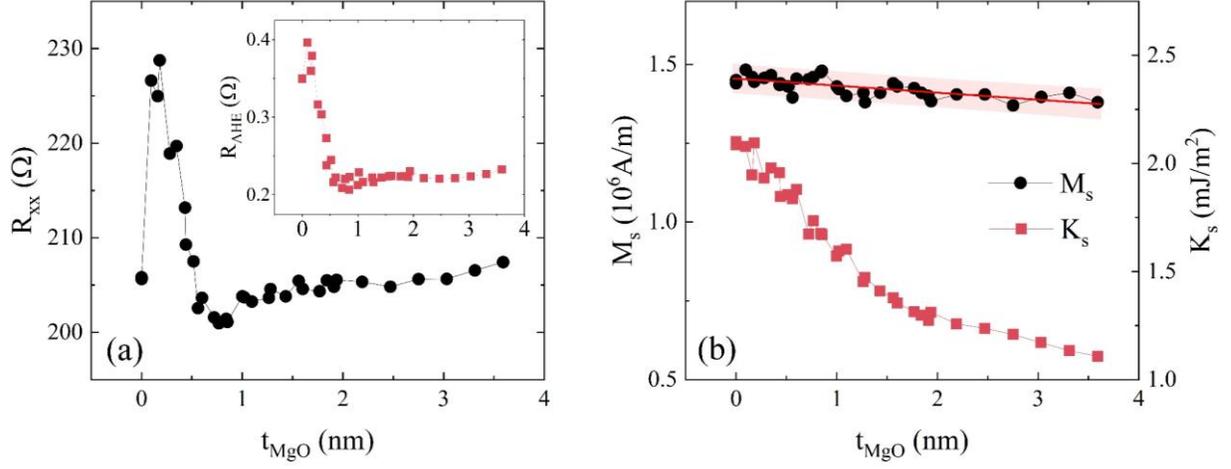

FIG. 6. (a) Longitudinal electrical resistance ($R_{xx}$), the anomalous Hall resistance ($R_{AHE}$), (b) saturation magnetization ($M_s$) and surface magnetic anisotropy ($K_s$) as a function of the MgO layer thickness ($t_{MgO}$). The straight line is a linear fit of the $M_s$ versus $t_{MgO}$.

Figure 6(b) shows the saturation magnetization ($M_s$) and the surface magnetic anisotropy ($K_s$) as a function of $t_{MgO}$. The $M_s$ shows a slight linear decrease with increasing $t_{MgO}$. This implies a soft oxidation of the Co layer as more oxygen becomes available for larger $t_{MgO}$. It corresponds to an oxidation of about 0.11 nm of Co for the largest $t_{MgO}$. This is also reflected in the evolution of the surface magnetic anisotropy, which shows a strong decrease with the thickness of the MgO layer. It is well known that the surface magnetic anisotropy at the Co/MgO interface is related to the hybridization between O $p$ and Co $d_{z^2}$ orbitals, and that it decreases strongly in the case of suboptimal (over – or under –) oxidation of the Co layer [51-53].

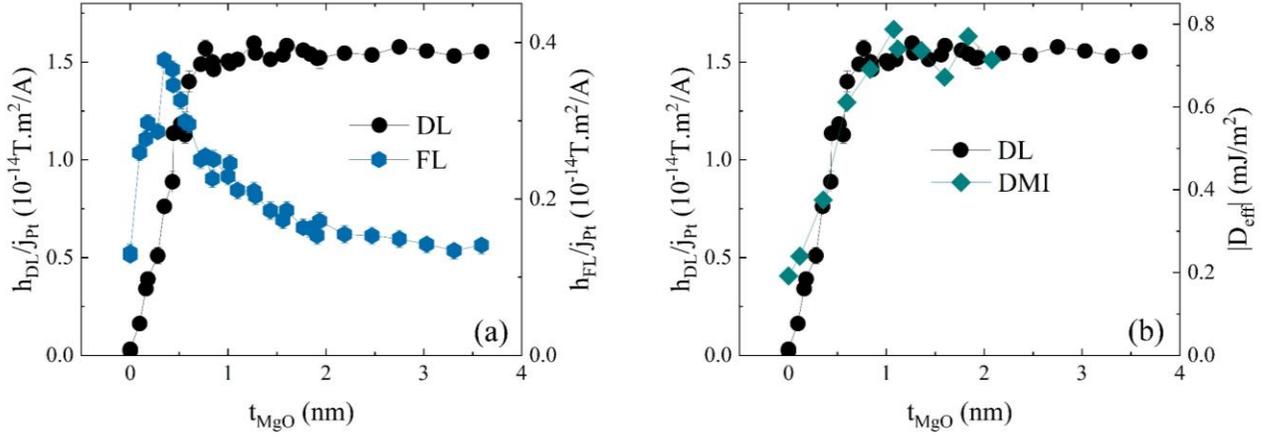

FIG. 7. (a) Damping-like ($h_{DL}/j_{Pt}$) and field-like ($h_{FL}/j_{Pt}$) effective fields normalized by the charge current density through the Pt layer ($j_{Pt}$) as a function of the MgO layer thickness ($t_{MgO}$). (b) The effective iDMI constant and $h_{DL}/j_{Pt}$, as a function of $t_{MgO}$.

Figure 7(a) shows the $h_{DL}/j_{Pt}$ and $h_{FL}/j_{Pt}$ as a function of $t_{MgO}$. The two effective fields $h_{DL}$ and $h_{FL}$ were normalized by the charge current density through the bottom Pt layer and were corrected by a factor that accounts for the slight decrease in $M_s$ with $t_{MgO}$, considering that the effective fields should scale inversely with the magnetic volume. Concerning $h_{DL}$, it starts from a negligible small value, shows an abrupt increase, and then remains relatively constant as a function of $t_{MgO}$. This behaviour can be understood by considering SHE in Pt as the main source of $h_{DL}$. In the case of the symmetric Pt/Co/Pt structure, the effects of SHE generated the spin currents from the top and bottom Pt layers cancel each other out, rendering $h_{DL}$ negligibly small. However, as $t_{MgO}$ is increased and the MgO layer becomes continuous, the spin current from the top Pt layer is blocked, and $h_{DL}$ is determined solely by the spin current from the bottom Pt layer, which does not depend on the MgO layer thickness. Thus, $h_{DL}$ remains independent of $t_{MgO}$ for values larger that 0.6-0.7 nm.

The behaviour of $h_{FL}$ is quite different. Unlike $h_{DL}$, in the case of the symmetric Pt/Co/Pt structure, the $h_{FL}$ is not negligible small. Moreover, after the first increase, $h_{FL}$ decreases with $t_{MgO}$, rather than remaining constant as $h_{DL}$. This indicates the existence of an additional mechanism at the Co/MgO interface influencing the FL-SOT.

The reduction of PIM in the top Pt layer with increasing $t_{MgO}$ could, in principle, affect $h_{FL}$. Nonetheless, even if this is the case, for the MgO layer thicknesses larger than 0.6-0.7 nm, for which the MgO layer is continuous, one would expect that the PIM in Pt to become negligible, since ultrathin interlayers are known to extinguish the PIM at the Pt/Co interface [49]. Because $h_{FL}$ is decreasing even above this thickness, the observed reduction cannot be attributed to a decrease of PIM.

Oxygen migration towards the bottom Pt/Co interface could also affect the FL-SOT. To exclude this possibility, we performed additional measurements of the interfacial Dzyaloshinskii-Moriya interaction (iDMI) using Brillouin light scattering spectroscopy [29]. iDMI is an interfacial interaction which is mainly given by the Pt/Co interface in Pt/Co/MgO structures [54]. In case of oxygen migration towards the bottom Pt/Co one would expect the DMI to be affected since the interaction is extremely sensitive to interfacial details [42]. Figure 7(a) shows the effective iDMI constant ($D_{\text{eff}}$) alongside with $h_{DL}/j_{Pt}$, as a function of $t_{MgO}$. Except for the symmetric Pt/Co/Pt structure with $t_{MgO} = 0$, the $D_{\text{eff}}$ and $h_{DL}/j_{Pt}$ follow the same trend and remain independent of $t_{MgO}$ for thicknesses larger than 0.7 nm. This indicates that for $t_{MgO}$ up to 0.7 nm the iDMI at the bottom Pt/Co interface and top Co/MgO/Pt interface adds destructively and the iDMI at the top Co/MgO/Pt interface decreases with increasing $t_{MgO}$. Furthermore, since for $t_{MgO}$ larger than 0.7 nm the $D_{\text{eff}}$ value is in agreement with the one expected for the Pt/Co interface [42,55], it follows that the iDMI at the top Co/MgO/Pt interface becomes negligibly small and the iDMI at the bottom Pt/Co interface is unaffected by increasing the MgO layer thickness.

Interestingly, for the symmetric Pt/Co/Pt structure both $D_{\text{eff}}$ and $h_{FL}/j_{Pt}$ have nonnegligible values. This can be understood by considering that both iDMI and FL-SOT are of interfacial nature and scale with iSOC. The bottom Pt/Co surface anisotropy was estimated to be $K_s^{Pt/Co} = 1.26 \pm 0.05$ mJ/m$^2$ (see Fig. S4 from [29]) by considering that only this interface contributes to $K_s$. At the same time, the surface anisotropy for the symmetric Pt/Co/Pt structure is $K_s^{Pt/Co/Pt} = 2.1 \pm 0.05$ mJ/m$^2$ [see Fig. 6(b)]. This shows that the top Co/Pt interface contributes a maximum of $K_s^{Co/Pt} = 0.84 \pm 0.1$ mJ/m$^2$ to the total surface magnetic anisotropy. The presence of larger interfacial anisotropy arising from the bottom Pt/Co interface than from the top Co/Pt one is a known feature in Pt/Co/Pt structures [56,57]. Given that the surface magnetic anisotropy at the Pt/Co and Co/Pt interfaces scales with iSOC, it can be inferred that the iSOC at the bottom Pt/Co interface is higher than at the top Co/Pt interface. Consequently, one expects that both iDMI and FL-SOT to be higher at the bottom Pt/Co interface than at the top Co/Pt interface. Therefore, although for the symmetric structure both iDMI and FL-SOT add destructively, they do not cancel out since they have different magnitudes at the two interfaces.

Our experimental observations indicate that the decrease of $h_{FL}$ for $t_{MgO}$ larger than 0.7 nm is not related to the bottom Pt/Co interface, but with the upper Co/MgO one. From Fig.8(a) one can see that $h_{DL}/j_{Pt}$ does not scale with $K_s$, which is expected considering that SHE in the bottom Pt layer is responsible for DL-SOT. However, from Fig. 8(b) one can observe that $h_{FL}/j_{Pt}$ and $K_s$ are well corelated, indicating that the same mechanism of interfacial oxidation affects both quantities. First principle calculations [53] showed that the presence of oxygen at the Co surface in Pt/Co structures leads to a transfer of electrons from Co to oxygen and, consequently, to a reduction of $K_s$, in line with our observations. The decrease of $M_s$ and $K_s$ with increasing $t_{MgO}$ points out to the increased Co oxidation and, consequently, to Co to oxygen charge transfer, as more oxygen becomes available with increased $t_{MgO}$. The Co/MgO interfacial charge transfer also affects the Rashba field at this interface. This in turn, will affect the FL-SOT generated at the same interface, since they are strongly corelated [16,48,58]. Our results clearly indicate that for $t_{MgO} > 0.7$ nm a FL-SOT develops at the top Co/MgO interface via REE, whose strength increases with $t_{MgO}$ and has an opposite sign than the one generated by the SHE or interfacial effects at bottom Pt/Co interface. We would also like to point out that for $t_{MgO} > 0.7$ nm, the $R_{xx}$ exhibits only a slight increase of approximately 2%. This small variation ensures that the current distribution through the stack remains relatively unchanged upon increasing $t_{MgO}$ from 0.6-0.7 nm up to 3.6 nm. Thus, the variations of $h_{FL}$ shown in Figures 7 and 8, will still hold even if we normalize the $h_{FL}$ by the current density through the Co layer. This furthermore supports our observation that the mechanism responsible for $h_{FL}$ at the Co/MgO interface is the REE.

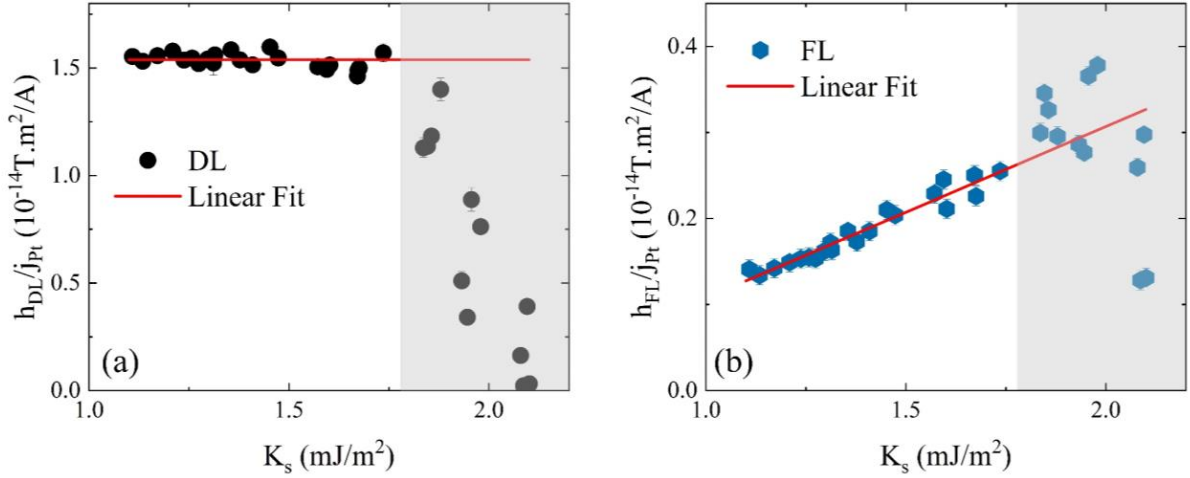

FIG. 8. (a) Damping-like ($h_{DL}/j_{Pt}$) and (b) field-like ($h_{FL}/j_{Pt}$) effective fields normalized by the charge current density through the Pt layer ($j_{Pt}$) as a function of surface magnetic anisotropy ($K_s$). The shaded areas correspond to MgO layer thickness lower than 0.72 nm. The lines are linear fits of the points outside the shaded areas.

*Conclusions*

In summary, we showed that in the case of the Pt/Co/MgO structures both bulk and interface effects influence the SOTs. We identified three mechanisms responsible for the SOTs acting on the FM layer magnetization: (i) SHE in Pt, (ii) REE and/or interfacial spin-currents induced by iSOC at the Pt/Co interface, and (iii) REE at the top Co/MgO interface. The behaviour of $h_{DL}$ with varying $t_{Pt}$ and $t_{Co}$ indicates that bulk SHE in Pt is mainly responsible for DL-SOT. The insertion of a Pd ultrathin layer at the Pt/Co interface leads to a decrease in $h_{DL}$, attributed to the variation of interfacial spin transparency. Additionally, the consistent insensitivity of $h_{DL}$ to variations in iSOC at the bottom Pt/Co interface and oxidation at the Co/MgO interface provides further support for the bulk SHE [mechanism (i)] as the primary source of DL-SOT. Conversely, the behaviour of $h_{FL}$ with varying $t_{Co}$ indicates that besides SHE other interfacial mechanisms contribute to FL-SOT. The significant decrease of $h_{FL}$ with decreasing iSOC at the Pt/Co interface points to the dominant interfacial nature of FL-SOT, driven by iSOC-induced interfacial generated spin currents or REE at the Pt/Co interface [mechanism (ii)]. Moreover, we show that an additional FL-SOT develops at the top Co/MgO interface with opposite sign as the one generated by the SHE or interfacial effects at bottom Pt/Co interface and whose strength increases with Co/MgO interfacial oxidation [mechanism (iii)]. Our experimental results elucidate the origins of SOTs in Pt/Co/MgO structures, resolving ongoing debates about their bulk versus interfacial origins.

*Acknowledgements*

This work was supported by a grant of the Romanian Ministry of Education and Research, CNCS - UEFISCDI, project number PN-III-P4-ID-PCE-2020-1853, within PNCDI III. IMM acknowledges funding for this work from the European Research Council (ERC) under the European Union's Horizon 2020 research and innovation programs: ERC-StG Smart Design (638653) and ERC-PoC SOFT (963928).

# Supplemental Material

# Bulk and interface spin-orbit torques in Pt/Co/MgO thin film structures

M.S. Gabor[1], M. Belmeguenai[2], I.M. Miron[3]

[1]*Center for Superconductivity, Spintronics and Surface Science, Physics and Chemistry Department, Technical University of Cluj-Napoca, Str. Memorandumului, 400114 Cluj-Napoca, Romania*
[2]*Université Sorbonne Paris Nord, LSPM, CNRS, UPR 3407, F-93430 Villetaneuse, France*
[3]*Université. Grenoble Alpes CNRS, CEA, Grenoble INP, SPINTEC, Grenoble, France*

S1. Electrical and magnetic properties of the Si/SiO$_2$//Ta (2)/Pt (1.5-5)/Co (2)/MgO (2)/Ta (1.5) stacks.

S2. Electrical and magnetic properties of the Si/SiO$_2$//Ta (2)/Pt (5)/Co (1.4-4)/MgO (2)/Ta (1.5) stacks.

S3. Electrical and magnetic properties of the Si/SiO$_2$//Ta (2)/Pt (5)/Pd (0-1.8)/Co (2)/MgO (2)/Ta (1.5) stacks.

S4. Oersted field estimation for the Si/SiO$_2$//Ta (2)/Pt (5)/Co (2)/MgO (0-3.6)/Pt (5) samples.

S5. Details on the interfacial Dzyaloshinskii–Moriya interaction measurements.

## S1. Electrical and magnetic properties of the Si/SiO$_2$//Ta (2)/Pt (1.5-5)/Co (2)/MgO (2)/Ta (1.5) stacks.

Figure S1 (a) shows the electrical resistance ($R_{xx}$) of the Si/SiO$_2$//Ta (2)/Pt (1.5-5)/Co (2)/MgO (2)/Ta (1.5) stacks as a function of the thickness of the Pt layer ($t_{Pt}$), illustrating the decrease of $R_{xx}$ with increasing $t_{Pt}$. The inset shows the inverse of the resistance ($1/R_{xx}$) as a function of $t_{Pt}$, fitted within the Fuchs-Sondheimer (FS) model [1,2], to determine the resistivity of Pt layer. The data point in the inset of Fig. S1(a) at $t_{Pt} = 0$ corresponds to the Ta (2)/Co (2)/MgO (2)/Ta (1.5) stack deposited in the same run. For such thicknesses, the Pt layer resistivity is not constant but increases with decreasing $t_{Pt}$ due to the enhancement of the diffusive interface scatterings [3]. In the parallel resistor model, the inverse resistance of the stack is given by $1/R_{xx} = 1/R_0 + 1/R_{Pt}$, where $R_0$ is the resistance of all the layers except the Pt layer, and $R_{Pt}$ is the resistance of the Pt layer. Within the FS model, $R_{Pt} = \rho_{xx}L/wt_{Pt}$, where $w$ is the width of the strip, $L$ is the distance between the voltage probes, $t_{Pt}$ is the thickness of the Pt layer, and $\rho_{xx} = \rho_{xx0}(1 + 3\lambda/8t_{Pt})$, with $\rho_{xx0}$ the bulk resistivity of the platinum layer and $\lambda$ the mean free path of the conduction electrons. The fit shown in the inset of Fig. S1(a) gives $\rho_{xx0} = 23\ \mu\Omega$cm and $\lambda = 13$ nm. The calculated and $\rho_{xx}$ is shown in Fig.S1(b). The current density through the Pt layer was determined as $j_{Pt} = IR_{xx}/\rho_{xx}L$, where $I$ is the total current through the. Using these values, the Oersted field due to the charge current passing through the Pt layer was calculated as $h_{Oe} = \mu_0 j_{Pt} t_{Pt}/2$.

The anomalous Hall resistance, $R_{AHE}$, was determined by applying an out-of-plane field and measuring the transverse voltage ($V_{xy}$) which is then divided by the electrical current ($I$) passing through the device to give the transverse resistance $R_{xy} = V_{xy}/I$. The $R_{AHE}$ is calculated as $[R_{xy}(+M_z) - R_{xy}(-M_z)]/2$, where $R_{xy}(+/-M_z)$ is the transverse resistance for positive/negative saturation. The inset of Fig. 1(c) from the main text shows a representative AHE resistance measurement. The increase of the $R_{AHE}$ relative to the Pt/Pd/Co samples (see Fig. S5) is ascribed to the Pt/Co interface. It is well known that interfacial spin-orbit coupling at the Pt/Co interfaces could induce a large AHE contribution with respect to bulk Co [4-6]. The further decrease of the $R_{AHE}$ with $t_{Pt}$ is due to the current shunting through the bulk of the Pt layer. The planar Hall resistance, $R_{PHE}$, follows the trend of $R_{AHE}$. This is expected since PHE is an anisotropic magnetoresistance effect which is influenced by the interfacial spin-orbit coupling at the Pt/Co interface.

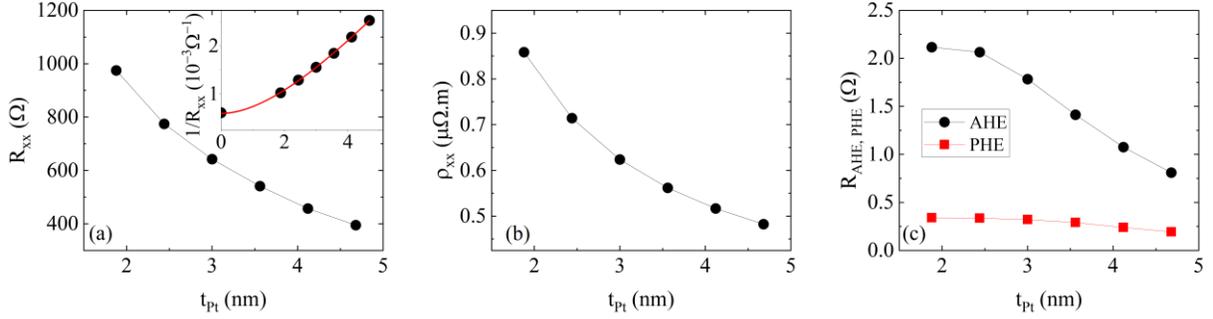

FIG. S1. (a) Electrical resistance of the stack, inset shows the inverse of the resistance fitted within the FS model (b) resistivity of the Pt layer, (c) anomalous and planar Hall resistances as a function of the Pt layer thickness.

The saturation magnetization ($M_s$) of the samples was measured at room temperature employing a vibrating sample magnetometer and is depicted in Fig. S2(a). The $M_s$ is roughly independent of $t_{Pt}$ and around $1.43 \times 10^6$ A/m. The anisotropy field ($\mu_0 H_k$) was determined from AHE measurements, as indicated in the inset of Fig. 1(c) from the main text. As seen in Fig. S2(b), it decreases with increasing $t_{Pt}$ and saturates for larger values. The relative low value of $\mu_0 H_k$ compared to the dipolar field $\mu_0 M s \approx$ 1.8 T is related to the presence surface magnetic anisotropy at the Pt/Co interface [7]. The $t_{Pt}$ dependence of the $\mu_0 H_k$ is most likely related to the evolution of the strains which are known to affect the perpendicular magnetic anisotropy in such systems [6, 8].

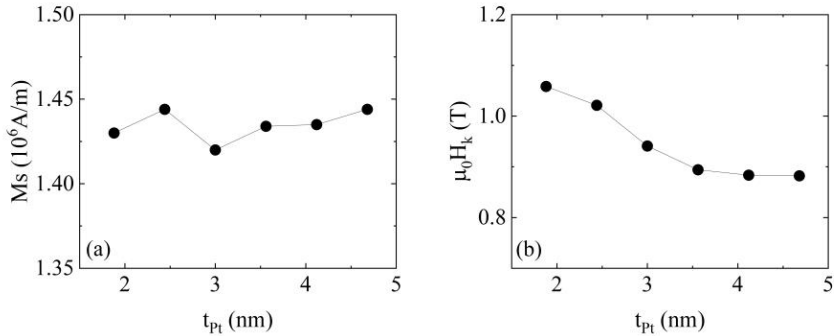

FIG. S2. (a) Saturation magnetization and (b) anisotropy field as a function of the Pt layer thickness.

## S2. Electrical and magnetic properties of the Si/SiO2//Ta (2)/Pt (5)/Co (1.4-4)/MgO (2)/Ta (1.5) stacks.

Figure S3 (a) shows the electrical resistance ($R_{xx}$) of the Si/SiO2//Ta (2)/Pt (5)/Co (1.4-4)/MgO (2)/Ta (1.5) stack as a function of the Co layer thickness ($t_{Co}$), which decreases with the increase of the thickness of the Co layer. The inset shows the inverse of the resistance as a function of $t_{Co}$, fitted within FS model to determine the resistivity of Co layer [Fig. S3(b)]. The data point in the inset of Fig. S3(a) at $t_{Co} = 0$ corresponds to the Ta (2)/Pt (5)/MgO (2)/Ta (1.5) stack deposited in the same run. The Co layer resistivity increases with decreasing $t_{Co}$ due to the enhancement of the diffusive interface scatterings. The current passing through the Co layer was determined as $I_{Co} = Iwt_{Co}/\rho_{xx}L$, where $I$ is the total current through the sample, $w$ is the width of the strip, $t_{Co}$ is the thickness of the Co layer, $\rho_{xx}$ is the Co layer resistivity shown in Fig. S3(b) and $L$ is the distance between the voltage probes. The current density passing through the Pt layer was calculate as $j_{Pt} = (I - I_{Co})/wt_{Pt}$. Here we assumed that the current shunting through the Ta (2) layer is negligible. This assumption is reasonable since the resistivity of the Ta (2) layer is more than one order of magnitude larger than that of the Pt (5) layer, which gives a current shunting less than 3% through the Ta (2) layer. Finally, the Oersted field was calculated as $h_{Oe} = \mu_0 j_{Pt} t_{Pt}/2$.

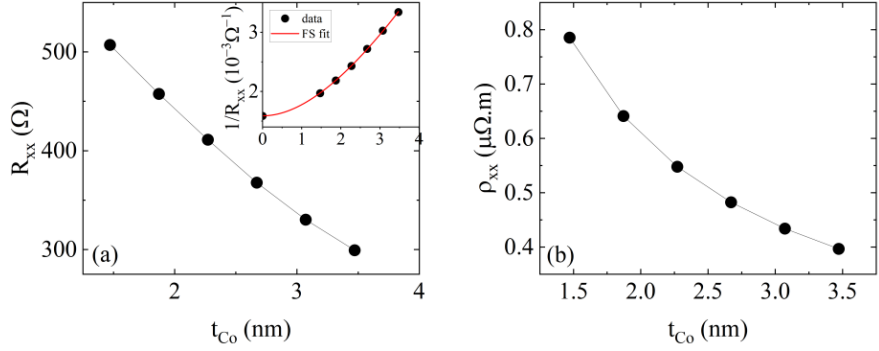

FIG. S3. (a) Electrical resistance of the stack. The inset shows the inverse of the resistance fitted within the FS model (b) resistivity of the Co layer.

The $M_s$ of the samples is depicted in Fig. S4(a), it is rather constant with a small decrease for lower $t_{Co}$. By fitting the product $M_s \times t_{Co}$ versus $t_{Co}$ we determined a relatively small magnetic dead layer, approximately 0.06-0.07 nm in thickness, which is likely attributed to a slight oxidation of the Co at the Co/MgO interface. The effective anisotropy constant was calculated as $K_{eff} = -1/2\mu_0 M_S H_K$. The magnetic anisotropy can be phenomenologically separated into a surface and a volume contribution using the relation [9] $K_{eff} \times t_{Co} = K_v \times t_{Co} + K_s$. Figure S4(b) shows the $K_{eff} \times t_{Co}$ as a function of $t_{Co}$ and the linear fit used to extract the volume and surface anisotropies. These values are in agreement with literature and the fact that the data does not deviate from the linear behaviour indicated that strains in Co are negligible for such thicknesses [8].

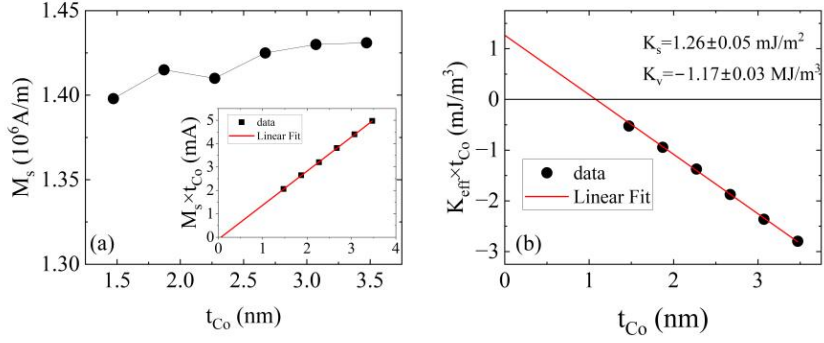

FIG. S4. (a) Saturation magnetization and (b) $K_{eff} \times t_{Co}$ as a function of $t_{Co}$ and the linear fit used to extract the volume and surface anisotropies.

## S3. Electrical and magnetic properties of the Si/SiO$_2$//Ta (2)/Pt (5)/Pd (0-1.8)/Co (2)/MgO (2)/Ta (1.5) stacks.

Figure S5 (a) shows the electrical resistance ($R_{xx}$) of the Si/SiO$_2$//Ta (2)/Pt (5)/Pd (0-1.8)/Co (2)/MgO (2)/Ta (1.5) stacks as a function of the total thickness of the Pt and Pd layers $t_{Pt+Pd}$. The insertion of the Pd layer leads to a linear variation of the inverse of the resistance ($1/R_{xx}$) with respect to $t_{Pt+Pd}$, as illustrated in the inset of Fig. S5(a). This indicates that the diffuse scatterings at the Pt/Pd interface are negligible. This is expected since Pt and Pd share similar crystal structures. Both belong to the Fm-3m space group, with closely matched lattice parameters ($a_{Pt} = 0.392$ nm and $a_{Pd} = 0.389$ nm), promoting high-quality layer-by-layer growth of Pd on Pt. Furthermore, their similar bulk electrical resistivities ($\rho_{Pt} = 106$ nΩm and $\rho_{Pt} = 105$ nΩm) corroborated with the negligible diffuse scatterings at the Pt/Pd interface ensure uniform current distribution in the Pt/Pd bilayer. In fact, this is one of the main reasons why we chose Pd as an interlayer. Our strategy was to use an interlayer that would not disturb the current flow through the structure and that has a lower spin-orbit coupling (SOC) than Pt.

The premise of uniform current flow through the Pt/Pd bilayer is further supported by analysing the linear relationship between $1/R_{xx}$ and $t_{Pt+Pd}$. A direct application of the FS model to the data presented in the inset of Fig. S5(a) would not yield meaningful results due to the lack of significant *curvature* in the data, leading to unphysical values and unreasonable uncertainties for the fitting parameters. Nonetheless, by performing a first-order Taylor expansion of the FS model around a thickness of $t_{Pt+Pd} = 6$ nm, we derive a linear dependence with a slope given by $\frac{32w(8+\lambda)}{\rho_{xx0}L(16+\lambda)^2}$, where $w$ is the width of the strip, $L$ is the distance between the voltage probes, $\rho_{xx0}$ the bulk resistivity of the Pt/Pd bilayer layer and $\lambda$ the mean free path of the conduction electrons in nm. Calculating this slope using the bulk resistivity $\rho_{xx0}$ and the mean free path $\lambda$ of the conduction electrons obtained in section S1 for the samples with Pt layer variable thickness, yields a value of $6.95 \times 10^5$ $[\Omega m]^{-1}$, which is in agreement with the one obtained by linear fitting the data in the inset of Fig. S5(a), $(6.89 \pm 0.08) \times 10^5$ $[\Omega m]^{-1}$. This indicates that the values of the bulk resistivity $\rho_{xx0}$ and the mean free path $\lambda$ of the conduction electrons obtained for the Pt layer are also relevant for the Pt/Pd bilayer and that our assumption of uniform current distribution in the Pt/Pd bilayer is correct.

The Oersted field was calculated in this case as $h_{Oe} = \mu_0 j_{Pt+Pd} t_{Pt+Pd}/2$, where $j_{Pt+Pd}$ is the current density through the Pt/Pd bilayer. The current density was calculated by assuming uniform current distribution in the Pt/Pd bilayer and by considering a parallel resistor model in which the decrease of the resistance with increasing the thickness of the Pt/Pd bilayer is due to the current shunting through the bilayer.

Figure S5 (b) shows the anomalous and planar Hall resistances as a function of $t_{Pt+Pd}$. Both resistances show a decrease with increasing $t_{Pt+Pd}$. The decrease of the AHE resistance goes beyond the current shunting through the increasing thickness of the Pt/Pd bilayer (responsible for about 32% drop), and is due to the decrease of the interfacial spin orbit coupling (iSOC) by the insertion of the Pd layer at the Pt/Co interface[4-6].

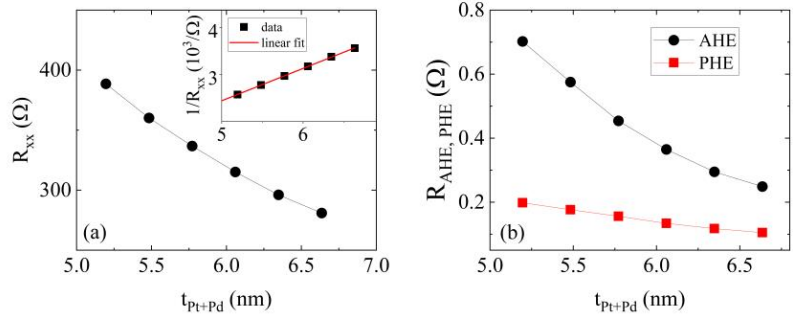

FIG. S5. (a) Electrical resistance of the stack, inset shows the inverse of the resistance (b) anomalous and planar Hall resistances as a function of the Pt+Pd layer thickness.

The $M_s$ of the samples is depicted in Fig. S6(a). It is rather independent of the thickness of the Pd layer. Thus, an average value of $1.437 \times 10^6$ A/m was used. The effective anisotropy was determined as $K_{eff} = -1/2\mu_0 M_S H_K$ and the surface magnetic anisotropy ($K_s$) was obtained from the relation $K_{eff} \times t_{Co} = K_v \times t_{Co} + K_s$, where $t_{Co} = 2$ nm. For $K_V$ we used the value determined for the samples with variable $t_{Co}$. Figure S6(b) shows the $K_s$ as a function of $t_{Pt+Pd}$, while the inset depicts the anisotropy field dependence on $t_{Pt+Pd}$.

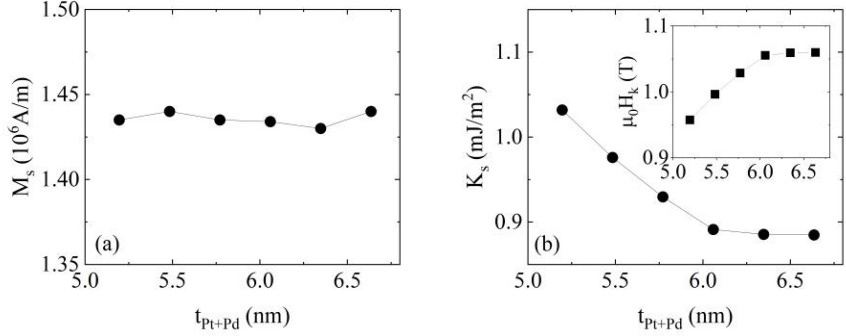

FIG. S6. (a) Saturation magnetization, (b) surface magnetic anisotropy and the anisotropy field (inset) as a function of the Pt+Pd layer thickness.

## S6. Oersted field estimation for the Si/SiO₂//Ta (2)/Pt (5)/Co (2)/MgO (0-3.6)/Pt (5) samples.

To precisely determine $h_{FL}$, it is necessary to estimate the Oersted field produced by the charge current. In the case of the Si/SiO₂//Ta (2)/Pt (5)/Co (2)/MgO (0-3.6)/Pt (5) structure, both the lower and upper Pt layers produce Oersted fields that add destructively (see Fig. S7). To accurately measure the electrical resistance of the top Pt layer, we employed a technique where two sample stacks were deposited in a single run, using the shutter as an *in-situ* mask. This approach allowed us to isolate and directly measure the resistance of the top Pt. The configuration of the first stack, Si/SiO₂//Ta (2)/Pt (5)/Co (2)/MgO (3.6) lacks the top Pt layer and its resistance is denoted as $R_{Pt(0)}$. The second configuration, Si/SiO₂//Ta (2)/Pt (5)/Co (2)/MgO (3.6)/Pt (5) includes the 5 nm Pt layer on top and its resistance is denoted $R_{Pt(5)}$. We used a parallel resistor model to determine the resistance of the top Pt layer as $R_{Pt\_top} = 1/(1/R_{Pt(5)} - 1/R_{Pt(0)})$, from which we were able to determine the resistivity of the top Pt layer. The electrical resistivity of the bottom Pt layer was determined in section S1. The electrical resistance of the top Pt layer was consistently slightly higher (around 5%) that the one of the bottom Pt layer. Using the values of the resistivities of both Pt layers and the resistance of the whole stack, we calculated the resistivity of the Co layer. We always checked that this calculated value to be consistent with the one obtained in sections S1 and S2. This approach provides a detailed insight into the electrical behaviour of the entire structure.

Since the electrical resistance of the Si/SiO$_2$//Ta (2)/Pt (5)/Co (2)/MgO (0-3.6)/Pt (5) stack is not strictly constant but depends on $t_{MgO}$ [see Fig. 6(a) from main text], we determined the Oersted field in two scenarios. (I) We considered that the increase in the resistance is due to the increase of the interface diffusive scattering at the Co/MgO and MgO/Pt interfaces. In this scenario, the resistances of the Co and top Pt layers increase relatively to account for the increase of the resistance of the whole stack. The Oersted ($h_{Oe}/j_{Pt} - I$) field calculated within this scenario as a function of $t_{MgO}$ is shown in Fig. S7, alongside with the measured transverse effective field ($h_T/j_{Pt} - I$) and the field-like effective field ($h_{FL}/j_{Pt} - I$). Of course, this scenario is more suitable for MgO thicknesses less than 0.7 nm, where the MgO layer is not perfectly continuous. (II) In the second scenario, we assume that the increase in resistance is a result of increased oxidation of the Co layer, effectively reducing its thickness. Consequently, the total resistance increase can be attributed to the Co layer alone. In this context, the total increase of the resistance is due only to the Co layer. The Oersted ($h_{Oe}/j_{Pt} - II$) field calculated within this scenario as a function of $t_{MgO}$ is shown in Fig. S7, along with the measured transverse effective field ($h_T/j_{Pt} - II$) and the field-like effective field ($h_{FL}/j_{Pt} - II$). This scenario is better suited for MgO thicknesses greater than 0.7 nm, where the MgO film is continuous. From Fig. S7 one can see that the Oersted field corrections are small in both scenarios, except for MgO thicknesses less than 0.7 nm. For larger thicknesses, the Oersted field corrections are roughly the same in both scenarios. Consequently, we have utilized scenario (I) to account for the Oersted field across the entire MgO thickness range.

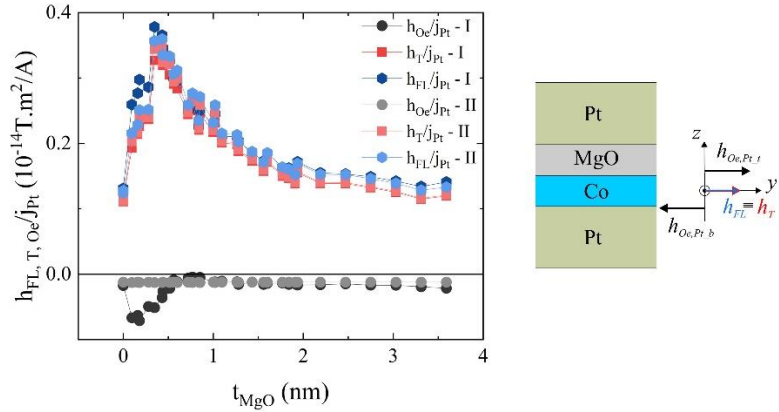

FIG. S7. Estimation of the Oersted field.

**S7. Details on the interfacial Dzyaloshinskii–Moriya interaction measurements.**

Brillouin light scattering (BLS) was employed to investigate the interfacial Dzyaloshinskii–Moriya interaction (iDMI) and the perpendicular magnetic anisotropy (PMA). This was achieved by measuring the frequency mismatch $\Delta F$ ($\Delta F = F_S - F_{aS}$) between the spin wave frequencies corresponding to the Stokes ($F_S$) and anti-Stokes ($F_{aS}$) lines. The variation of the $\Delta F$ versus $k_{SW}$ was utilized to characterize the strength of the iDMI from the relation $\Delta F = D_{eff} \frac{4\gamma}{2\pi M_s} k_{SW}$ [10], where $D_{eff}$ represents the effective DMI constant, characterizing the iDMI strength, and $\gamma$ is the gyromagnetic ratio. Figure S8 shows Representative measured BLS spectra and the variation of the frequency mismatch $\Delta F$ with $k_{SW}$ for samples with various MgO layer thicknesses, used to extract the effective DMI constant.

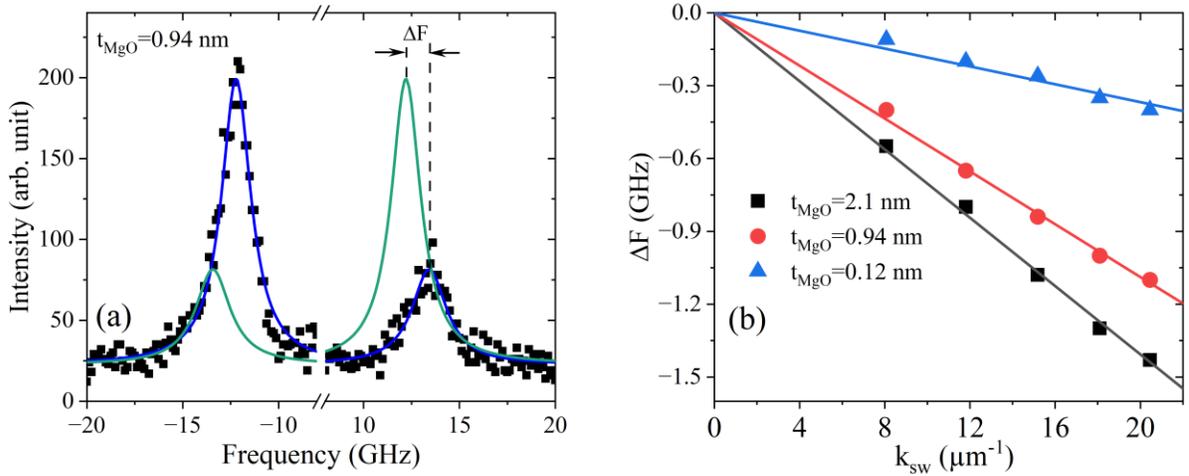

FIG. S8. (a) Representative measured BLS spectra. Symbols refer to experimental data and solid lines are Lorentzian fits. Fits corresponding to negative applied fields (green lines) are presented for clarity and direct comparison of Stokes and anti-Stokes frequencies. (b) Variation of the frequency mismatch $\Delta F$ with $k_{SW}$ for samples with various MgO layer thicknesses.